\documentclass[prl,aps,twocolumn,showpacs,preprintnumbers,amsmath,amssymb]{revtex4}
\usepackage{mathrsfs}
\usepackage{amsmath,amssymb,graphicx}
\usepackage {amssymb}
\usepackage{color}
\newcommand{\nc}{\newcommand}
\nc{\bea}{\begin{eqnarray}} \nc{\eea}{\end{eqnarray}}
\nc{\be}{\begin{equation}} \nc{\ee}{\end{equation}}

\newcommand\s{\sigma}

\nc{\ga}{\gamma} \nc{\x}{{\bf x }} \nc{\kk}{{\bf k }} \nc{\f}{{\bf f
}} \nc{\T}{ \theta (s_i (t)- \s) } \nc{\TT}{ \theta (s_i (t_{ r \, i
} )- \s) } \nc{\br}{   (s_i (t)- \s)  } \nc{\fa}{\phi_1}
\nc{\fb}{\phi_2}

\input{epsf.sty}

\begin{document}

\markright{CYCU-HEP-11-06}

\title{Higgs Inflation in Horava-Lifshitz Gravity}

\author{Taotao Qiu$^1$\footnote{xsjqiu@gmail.com} and Debaprasad Maity$^{2,3}$\footnote{debu.imsc@gmail.com}}

\affiliation{$^1$ Department of Physics, Chung-Yuan Christian
University, Chung-li 320, Taiwan\\
$^2$ Department of Physics and Center for Theoretical Sciences,
National Taiwan University, Taipei 10617, Taiwan\\
$^3$ Leung Center for Cosmology and Particle Astrophysics National
Taiwan University, Taipei 106, Taiwan}


\begin{abstract}
We study the possibility of standard model Higgs boson acting as an
inflaton field in the framework of Horava-Lifshitz Gravity. Under
this framework, we showed that it is possible for the Higgs field to
produce right amount of inflation and generate scale invariant power
spectrum with the correct experimental value. Thanks to the
foliation preserving diffeomorphism and anisotropic space-time
scaling, it essentially helps us to construct this model without the
pre-existing inconsistency coming from cosmological and particle
physics constraints. We do not need to introduce any non-minimal or
higher derivative coupling term in an arbitrary basis either.

\end{abstract}

\maketitle

{\it Introduction.} It has been widely accepted that our universe
has experienced an exponential expansion in the very early stage of
its evolution. Thanks to this exponential expansion, it explains the
flatness and homogeneity of our universe. It also helps to get rid
of heavy particles such as monopoles which would otherwise alter the
present state of evolution of our universe. This exponential
expansion is called {\it Inflation}
\cite{Guth:1980zm,Albrecht:1982wi,Linde:1983gd}. Most interestingly,
inflation also helps the quantum fluctuations to evolve into a
classical curvature perturbation which eventually sources the seed
of structure formation in our universe. By choosing a particular
model of inflation, the primordial power spectrum for the curvature
perturbation can be made scale-invariant which fits very well with
the latest WMAP data \cite{Larson:2010gs}. Moreover, all the other
known cosmological observations also support the inflation in the
early universe.

Constructing a model of inflation has been the subject of interest
for the last several decades. The simplest and phenomenologically
most successful model of inflation so far is a model of a single
scalar field called inflaton which drives the exponential expansion.
One of the fundamental issues with the standard inflationary model
is the origin of the scalar field. As we know standard model of
particle physics which is most successful, contains a natural scalar
field called Higgs. Although Higgs field has not been observed yet,
it would be natural and also economical if one can identify this
Higgs as an inflaton field. However, because of the strong
self-coupling which is constrained by the particle physics, it has
been ignored in the past in the inflationary model building. The
reason can be seen from a straightforward estimation of the power
spectrum for a minimally coupled Higgs field with the following
action,
\be {\cal S}_H=\int
d^4x\sqrt{-g}[\kappa^2R-\frac{1}{2}D_\mu{\bf H}^\dag D^\mu{\bf
H}-\frac{\lambda_H}{4}({\bf H}^\dag{\bf H}-v^2)^2]~, \ee
where
$\kappa^2\equiv 1/(16\pi G)$, $R$ is the Ricci scalar, $\bf{H}$ is
Higgs boson doublet, $D_\mu$ is the covariant derivative with
respect to $SU(2)\times U(1)$ gauge group, $\lambda_H$ is the
self-coupling coefficient and $v$ is the vacuum expectation value
(VEV) of the Higgs. Hereafter, we adopt the metric to be
\be\label{metricfrw} ds^2=dt^2-a^2(t)dx^2~.\ee
At the energy scale
of inflation we can ignore the VEV of the Higgs field. The action
therefore turns out to be: \be {\cal S}_H=\int
d^4x\sqrt{-g}[\kappa^2R-\frac{1}{2}
\partial_\mu\phi\partial^\mu\phi-\frac{\lambda_H}{4}\phi^4]~,
\ee where we consider a real scalar field $\phi$ in place of ${\bf
H}$ for simplicity.  The equations of motion for the metric and
$\phi$ are: \be H^2=\frac{8\pi
G}{3}(\frac{1}{2}\dot\phi^2+\frac{\lambda_H}{4}\phi^4)~,~~~\ddot\phi+3H\dot\phi+\lambda_H\phi^3=0~,\ee
where $H$ denotes the Hubble parameter and ${\dot H} = dH/dt$. In
order to get sufficient efolding, we impose a ``slow-roll" condition
\be \epsilon\equiv-\frac{\dot
H}{H^2}\simeq\frac{\kappa^2}{\phi^2}\ll 1~,\ee which essentially
sets $\phi \gg 1$ in Planck unit. Now following Ref.
\cite{Stewart:1993bc}, the primordial power spectrum of the
curvature perturbation turns out to be: \be\label{spectrum} {\cal
P}_\zeta\equiv\frac{k^3}{2\pi}|\zeta|^2\sim\frac{H^4}{2\pi\dot\phi}\sim\frac{H^2}{\epsilon}\sim\frac{V^3}{V_\phi^2}\sim\lambda_H\phi^6~.\ee
If we take $\phi > 1$, the observed power spectrum ${\cal
P}_\zeta\sim 10^{-9}$ sets the limit on $\lambda_H$ to be $\leq
10^{-9}$. This is in direct conflict with the standard model
prediction of Higgs coupling $0.11<\lambda_H\lesssim0.27$
\cite{Amsler:2008zzb}. This severe constraint makes it difficult to
construct a minimally coupled Higgs inflationary model.

In order to get rid of this inconsistency, only recently people have
come up with a non-trivial modification of Higgs action with the
gravity
\cite{CervantesCota:1995tz,Bezrukov:2007ep,Germani:2010gm,Kamada:2010qe}.
The simplest non-minimal coupling term that has been introduced
\cite{Bezrukov:2007ep} is $\xi R\phi^2$ where $\xi$ is the coupling
constant. In this model considering the slow-roll parameter
$\epsilon\simeq \xi^{-2}\phi^{-4}$, the expression for the efolding
number becomes ${\cal N}\simeq\xi\phi^2$. Thus required amount of
${\cal N}$ gives $\phi\sim 10/\sqrt{\xi}$. Considering this
constraint on $\phi$ and $0.11<\lambda_H\lesssim0.27$, one can easily
produce the experimental value of the power spectrum ${\cal
P}_\zeta\sim\lambda_H\phi^4\sim\lambda_H\xi^{-2} \sim {\cal
O}(10^{-9})$ by choosing the new parameter $\xi>10^4$.
However, later it has been pointed out that, this scenario is plagued
with the unitarity problem. More specifically, at the quantum level,
this non-minimal coupling term $\xi
\delta\phi^2\partial^2\gamma $ ($\gamma\equiv Tr(\gamma_{\mu\nu})$),
where $\delta \phi = \phi -\phi_0$ and $\gamma_{\mu\nu} =
g_{\mu\nu}-\eta_{\mu\nu}$ are the quantum fluctuation around the
background \cite{Burgess:2009ea}, will violate the unitarity of S-matrix
at an energy scale $\Lambda\simeq \xi^{-1}$. This should be
considered as a cut-off for the effective theory. This scale scale turns
out to be much
below the typical fluctuation of the Higgs field during inflation as
discussed above \cite{Lerner:2009na}.

To circumvent the above mentioned problem, the authors in
\cite{Germani:2010gm} introduced an alternative kinetic coupling of
the Higgs field with gravity of the form
$G^{\mu\nu}\partial_\mu\phi\partial_\nu\phi$, where $G^{\mu\nu}$ is
the Einstein tensor. This new non-minimal coupling term also gives
rise to a unitarity bound $\Lambda(H)\simeq(2H^2/\kappa)^{1/3}$ but
the claim is that this bound is well above the gravitational energy
scale during inflation. However, soon after this, a careful analysis
has been done in \cite{Atkins:2010yg} and showed that unitariry is
actually violated in this model as well. In an another attempt
authors of \cite{Kamada:2010qe} have introduced a non-trivial higher
derivative kinetic term $G((\partial\phi)^2,\phi)\Box\phi$ in
addition to the usual Higgs Lagrangian. This construction is
inspired by the recently proposed theory called Galileon theory
\cite{Nicolis:2008in}. The important property of this new terms is
that it does not lead to an extra degrees of freedom (ghost) because
the equation of motion for the Higgs field is still second order in
derivative. This new term modifies the dispersion relation of the
scalar field and helps to produce the sufficient number of efolding
as well as required amplitude of power spectrum. However, more
detail study needs to be done regarding the unitarity problems of
this kind of model.

In this Letter, we propose a new scenario where Higgs inflation can
be realized in the framework of Horava-Lifshitz (HL) gravity
\cite{Horava:2009uw}. HL theory of gravity is known to be invariant
under a foliation perserving diffeomorphism \be
\tilde{x}^i=\tilde{x}^i(x^j,t)~,~~~\tilde{t}=\tilde{t}(t)~.\ee
Interestingly, the theory can be made power counting renormalizable
in four dimension if one introduces an anisotropic scaling
transformation of space and time like \be
\overrightarrow{x}\rightarrow b\overrightarrow{x}~,~~~t\rightarrow
b^3 t . \ee Moreover, it is also argued that in the low energy
limit, theory flows to the standard General Relativity (GR) where
the full diffeomorphism invariance is recovered as an emerging
symmetry. All these interesting properties trigger a spate of
research works in the diverse directions for the last few years, see
the current status of HL theory from the reviews
\cite{Mukohyama:2010xz}. Although the original version of Horava
gravity may be plagued by the extra unwanted degrees of freedom
\cite{Charmousis:2009tc}, later on different extensions have been
proposed in order to cure this \cite{Blas:2009qj,Horava:2010zj}. In
this letter we will adopt the original version of the Horava gravity
to construct the Higgs inflationary model, while leaving concerns
over the other versions of Horava gravity for our future study.

{\it Higgs Inflation in HL Gravity.} With the symmetry under
consideration, it is customary to consider the $3+1$ decomposition
of the space-time metric: \be\label{metricadm}
ds^2=N^2dt-h_{ij}(dx^i+N^idt)(dx^j+N^jdt)~,\ee where
$N(t,\overrightarrow{x})$, $N_i(t,\overrightarrow{x})$ and $h_{ij}$
are the lapse function and the shift vector and the spatial metric
respectively. The most general HL action without the condition of
detailed balance will be of the form \footnote{if we consider
healthy extension of Horava Gravity, the extra degree of freedom
will become dynamical and may give rise to isocurvature perturbation
and non-conserved curvature perturbation, breaking the scale
invariance of the spectrum. We will leave this case for future
work.}: \be\label{action} \mathcal{S}=\kappa^{2}\int
dtd^{3}xN\sqrt{g}(\mathcal{L}_{K}-\mathcal{L}_{V}+
\kappa^{-2}\mathcal{L}_{M})~,\ee
where \bea \mathcal{L}_{K}&=&K_{ij}K^{ij}-\lambda K^{2}~,\\
\mathcal{L}_{V}&=&2\Lambda-R+\kappa^{-2}(g_{2}R^{2}+g_{3}R_{ij}R^{ij})\nonumber\\
&+&\kappa^{-4}(g_{4}R^{3}+g_{5}RR_{ij}R^{ij}+g_{6}R_{j}^{i}R_{k}^{j}R_{i}^{k})\nonumber\\
&+&\kappa^{-4}(g_{7}R\nabla^{2}R+g_{8}(\nabla_{i}R_{jk})(\nabla^{i}R^{jk}))~,\\
\mathcal{L}_{M}&=&\frac{1}{2N^{2}}(\dot{\phi}-N^{i}\nabla_{i}\phi)^{2}-\mathcal{V}(\phi,g_{ij})~,\eea
here
$K_{ij}\equiv(\dot{h}_{ij}-\nabla_{i}N_{j}-\nabla_{j}N_{i})/(2N)$ is
the extrinsic curvature, and $\lambda$ is a free parameter. In IR
region, $\lambda$ flows to unity to recover GR.
 The potential term reads: \bea
\mathcal{V}(\phi,g_{ij})&=&V_{0}(\phi)+V_{1}(\phi)(\nabla\phi)^{2}+V_{2}(\phi)(\Delta\phi)^{2}\nonumber\\
&+&V_{3}(\phi)(\Delta\phi)^{3}+V_{4}(\phi)(\Delta^{2}\phi)\nonumber\\
&+&V_{5}(\phi)(\nabla\phi)^{2}(\Delta^{2}\phi)+V_{6}(\phi)(\Delta\phi)(\Delta^{2}\phi)~.\eea
Note that we consider $\phi$ to be the Higgs field, so the
background potential will be $V_0(\phi)=\lambda_H\phi^4/4$. The
background equations of motion for the metric (\ref{metricfrw}) and
field $\phi$ are: \bea \label{einstein1} &&
\kappa^{-2}(\frac{1}{2}\dot{\phi}^{2}+V_{0}(\phi))+3H^{2}(1-3\lambda)=0~,\\
\label{einstein2}&&\kappa^{-2}(\frac{1}{2}
\dot{\phi}^{2}-V_{0}(\phi))-3H^{2}(1-3\lambda)=2(1-3\lambda)\dot{H}~\\
&&\ddot\phi+3H\dot\phi+V_{0\phi}=0~,
\eea
where
$V_{0\phi}\equiv\partial V_0(\phi)/\partial\phi$.
We also set cosmological constant $\Lambda =0$. The
expression for the slow-roll parameter $\epsilon$ and the number of
efoldings ${\cal N}$ will be of the same form as that of the
standard GR, namely, \be\label{epsilonN} \epsilon\equiv-\frac{\dot
H}{H^2}\simeq\kappa^2\frac{V_{0\phi}^2}{V^2}~,~~~{\cal
N}\equiv\int_{t_i}^{t_f}Hdt\simeq\int_{\phi_i}^{\phi_f}
\frac{V}{V_{0\phi}}d\phi~.\ee

\begin{figure}[htbp]
\includegraphics[scale=0.4]{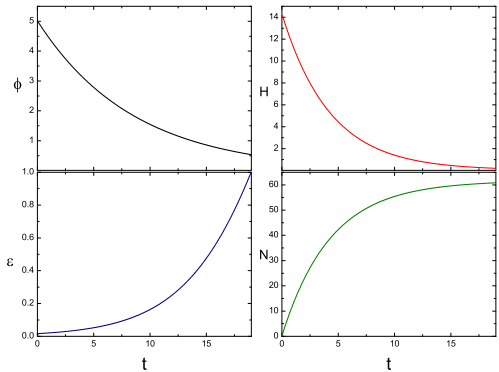}
\caption{(Colored online.) The evolution of $\phi$, $H$, $\epsilon$
and ${\cal N}$. Horizontal
axis is the cosmic time $t$. Parameters and initial values:
$\lambda=1.2$, $\lambda_H=0.2$, $v=0$, $\phi_i=5M_{pl}$,
$\dot\phi_i=0$. The normalization is $M_{pl}=1$. From the figures we
can see that at the end of inflation we have approximately
$\phi_f\simeq 0.54M_{pl}$ and $H_f\simeq 0.2M_{pl}$.}\label{all}
\end{figure}

As we have discussed, numerically it is easy to find out the
solution for slow-roll inflation for sufficiently long period as
shown in Fig. \ref{all}. We want to emphasize here that in usual
Higgs inflation scenario, the Hubble parameter (or scalar potential)
is very large because of large $\lambda_H$, which eventually leads
to a large curvature perturbation compared to the observed power
spectrum. Thanks to the foliation preserving diffeomorphism and
anisotropic space-time scaling of HL Gravity, in the UV limit it
turns out that the evolution of the curvature perturbation depends
only on the higher derivative term of the Higgs field and not on its
potential \footnote{The dependence of power spectrum on fixed energy
scale has been discussed in \cite{Mukohyama:2009gg}. However, in
their case the scalar field is curvaton, so the metric perturbation
is neglected and curvature perturbation was produced via curvaton
mechanisms.}. This is the key point that makes the Higgs inflation
feasible in the framework of HL gravity.

{\it Perturbations and Scale-Invariant Power Spectrum.} The
cosmological perturbation in the HL gravity has been widely studied
\cite{Gao:2009bx}. We expand the scalar field and spatial metric as
follows: \be\label{perturbation}
\phi(t,\overrightarrow{x})=\phi_0(t)+Q(t,\overrightarrow{x})~,~~~h_{ij}=a^2(t)e^{2\gamma(t,\overrightarrow{x})}\delta_{ij}~.\ee
In the cosmological perturbation theory, it is customary to write
down the equations of motion for the perturbation in terms of a
gauge invariant variable
 \be\label{curvature}
\zeta=\gamma-\frac{H}{\dot{\phi}_{0}}Q~,\ee which is a linear
combination of metric and scalar field perturbation. The
equation of motion for the gauge invariant perturbation $\zeta$ can
be further simplified by defining an another variable $u\equiv
a\sqrt{\mathcal{K}}\zeta$ where ${\cal K}$ is given in the appendix.
The final equation of motion of our interest would take a
 very simple form: \be\label{eom}
u^{\prime\prime}+\omega_{u}^{2}u=0~.\ee The modified dispersion
relation is \be
\omega_{u}^{2}=\frac{a^{2}\mathcal{M}^{2}}{\mathcal{K}}-(\mathcal{H}+\frac{\mathcal{K}^{\prime}}{2\mathcal{K}})^{2}-(\mathcal{H}+\frac{\mathcal{K}^{\prime}}{2\mathcal{K}})^{\prime}~,\ee
where ``prime" denotes the derivative w.r.t. conformal time $\eta$
with $dt=a(t)d\eta$. The expression for the effective mass
$\mathcal{M}$ is also given in the appendix.

It is intuitively obvious that the terms coming from higher spatial
derivative will be dominant in the expression for $\omega_{u}$ in UV
regime. In Fourier space, the leading order behavior of
$\omega_{u}^{2}$ would be $\omega_{u}^{2}\rightarrow
a^{2}\kappa^{-4}(1-\lambda)(3g_{8}-8g_{7})\bar{k}^{6}/(1-3\lambda)$,
where $\bar{k}\equiv k/a(t)$. $k$ is the wavenumber of fluctuation.
The equation of motion therefore becomes \be
u^{\prime\prime}+\frac{a^{2}(1-\lambda)(3g_{8}-8g_{7})\bar{k}^{6}}{\kappa^{4}(1-3\lambda)}u=0~.\ee
The solution turns out to be: \be
u\simeq\frac{1}{\sqrt{2\omega_u}}\exp\left(i\int\omega_ud\eta\right)~.\ee
Moreover, amplitude of the fluctuation freezes out at horizon
crossing where $\omega_u$ is comparable with the Hubble parameter
$H$. From the definition of power spectrum (\ref{spectrum}), we
find: \be {\cal P}_\zeta=\frac{k^3}{2\pi}\big|\frac{u}{a\sqrt{\cal
K}}\big|^2\simeq\kappa\left(\frac{1-\lambda}{4(1-3\lambda)(8g_7-3g_8)}\right)^{\frac{1}{4}}~.\ee
which is almost scale-invariant on the superhorizon scale. Note that
$\lambda$, $g_7$ and $g_8$ are all free parameters in our model,
with $\lambda$ either greater than 1 or smaller than $1/3$ in order
not to cause the ghost instabilities. Therefore, by choosing the
appropriate values of those parameters we can set
$(1-\lambda)/(8g_7-3g_8)(1-3\lambda)\sim{\cal O}(10^{-36})$, in
order to get $|{\cal P}_\zeta|\sim 10^{-9}$. We also would like to
emphasize here that because of no nonminimal coupling term in our
Lagrangian, we do no need to worry about the unitarity problems.

In the low energy regime, lower spatial derivatives terms in the
Lagrangian will start to dominate in the expression for
$\omega_{u}$. We can therefore approximate the expressions for
$\mathcal{K}$ and $\omega_{u}$ up to ${\cal O}(\bar{k}^2)$ as
follows \be \mathcal{K}\rightarrow
2\epsilon~,~~~\omega_{u}^{2}\rightarrow
k^{2}-\frac{(a\sqrt{\epsilon})^{\prime\prime}}{a\sqrt{\epsilon}}~,\ee
where $\epsilon$ is defined in (\ref{epsilonN}). One can easily see
that the eq. (\ref{eom}) reduces to the usual form of canonical
single field inflation in GR \footnote{The transfer from UV regime
to IR regime occurs when
$H_t\simeq[(1-\lambda)(8g_7-3g_8)/(1-3\lambda)]^\frac{1}{8}\sim{\cal
O}(10^{-4})$ \cite{Mukohyama:2009gg}. In our case $H_f\simeq
0.2M_{pl}$ is much larger than that. So in our model the inflation
happens well within the UV regime.}.




{\it End of the Inflation and Estimation of Reheating Temperature.}
As is pointed out in \cite{Bezrukov:2007ep}, one can assume that the
reheating happens right after the inflation ends due to the strong
interactions of the Higgs boson with the standard model particles.
At the end of the inflation one has $\epsilon \equiv -{\dot H}/H^2
\simeq 1$ which essentially sets the kinetic energy of the Higgs to
be of the same order as its potential energy i.e.
$\dot\phi_f^2\simeq V(\phi_f)=\lambda_H\phi_f^4/4$. By using
equations (\ref{einstein1}) and (\ref{einstein2}), this eventually
fixes the energy density of the scalar field at the time of
reheating as \be \rho_{\phi_f}=\frac{1}{2}\dot\phi_f^2+V(\phi_f)
\simeq\frac{3\lambda_H}{8}\phi_f^4~,\ee where numerical calculation
gives $\phi_f\simeq 0.54M_{pl}$. In thermal equilibrium the energy
density of the radiation field can be written as  \be
\rho_\gamma=\frac{g_*\pi^2T^4}{30}~,\ee where $g_*\simeq 106.75$ is
the numbers of relativistic degree of freedom and $T$ is the
equilibrium temperature. The reheating temperature $T_{reh}$ can
therefore be computed by assuming $\rho_\gamma\simeq\rho_{\phi_f}$
at the end of inflation. So, we get \be
T_{reh}\simeq(\frac{90\lambda_H\phi_f^4}{8g_*\pi^2})^\frac{1}{4}\sim
0.1M_{pl}~.\ee This is consistent with the constraint from Big Bang
Nucleosythesis.

{\it Discussions and Conclusion.} In this Letter we discussed about
the possibility of realizing Higgs inflation in the framework of
Horava-Lifshitz Gravity. One of the main problems with the usual
Higgs inflationary model in standard GR is that it produces a large
curvature perturbation because of large self-coupling. In order to
solve this problem various non-minimal coupling prescriptions of
Higgs with the gravity have been proposed. Most of these models are
not well established yet. In some models
\cite{Bezrukov:2007ep,Germani:2010gm} people have already found the
unitarity violation which makes those models inapplicable at
the inflationary energy scale. In this letter, we proposed a new way
of realizing Higgs inflation in the framework of HL theory. This
theory is invariant under the foliation preserving diffeomorphsim.
The space and time transforms differently under the scaling
transformation. As we have argued because of these different
space-time transformation behavior, the dynamics of the curvature
perturbation becomes independent of the Higgs potential in the high
energy limit, which eventually breaks the strong inter-connection
between the flatness of the scalar potential and the scale invariant
power spectrum. This in turn makes the Higgs inflation to work.
Furthermore, we estimate the reheating temperature and find it being
well within BBN constraints.

Connections between cosmology and particle physics is an important
arena of physics for the last several decades. Due to its novel
properties in the UV regime, Horava-Lifshitz theory may play an
important role in connecting the cosmology and particle physics. In
this Letter we tried to make a connection between these two through
Higgs inflation in the framework of HL gravity. However, we only
considered the scalar perturbations to leading order, while higher
order perturbations, such as non-Gaussianities in curvature
perturbation and corrections from loop-level Higgs scattering, are
also interesting. Furthermore, studying tensor perturbations in this
scenario are also important to fit the data. We leave all these
subjects to our future study.

{\it Acknowledgements.} T.Q. thanks X. Gao, S. Mukohyama, Y. S. Piao
and R. Brandenberger for useful discussions. The work at CYCU is
funded in parts by the National Science Council of R.O.C. under
Grant No. NSC99-2112-M-033-005-MY3 and No. NSC99-2811-M-033-008 and
by the National Center for Theoretical Sciences.

{\it Appendix.} The coefficients ${\cal K}$ and ${\cal M}$ that
appeared in the text are defined as: \be \mathcal{K}\equiv
c_{\gamma}+\frac{\Sigma_{1}^{2}}{4\omega_{\phi}^{2}}~,\ee \be
\mathcal{M}^{2}\equiv
m_{\gamma}^{2}-\frac{\Sigma_{2}^{2}}{4\omega_{\phi}^{2}}+\frac{1}{2}(\dot{f}_{\gamma}+3Hf_{\gamma})-\frac{1}{4a^{3}}\partial_{t}(a^{3}\frac{\Sigma_{1}\Sigma_{2}}{\omega_{\phi}^{2}})~.\ee
Furthermore, \be
\Sigma_{1}\equiv\tilde{f}_{\phi\gamma}+2c_{\gamma}(\dot{\phi}_{0}^{-1}H)^{\cdot}-f_{\phi\gamma}~,\ee
\be \Sigma_{2}\equiv
-2m_{\gamma}^{2}\frac{H}{\dot{\phi}_{0}}-m_{\phi\gamma}^{2}-(\dot{f}_{\phi\gamma}+3Hf_{\phi\gamma})-\dot{f}_{\gamma}\frac{H}{\dot{\phi}_{0}}-3f_{\gamma}\frac{H^{2}}{\dot{\phi}_{0}}~,\ee
\bea \omega_{\phi}^{2}&\equiv&
m_{\gamma}^{2}\frac{H^{2}}{\dot{\phi}_{0}^{2}}+m_{\phi\gamma}^{2}\frac{H}{\dot{\phi}_{0}}+m_{\phi}^{2}-c_{\gamma}(\dot{\phi}_{0}^{-1}H)^{\cdot2}-f_{\gamma}\frac{H}{\dot{\phi}_{0}}(\dot{\phi}_{0}^{-1}H)^{\cdot}\nonumber\\
&-&\tilde{f}_{\phi\gamma}(\dot{\phi}_{0}^{-1}H)^{\cdot}+\frac{1}{2}(3H\bar{f}_{\phi}+\dot{\bar{f}}_{\phi})~,\eea
\be \bar{f}_{\phi}\equiv
2c_{\gamma}(\dot{\phi}_{0}^{-1}H)^{\cdot}\frac{H}{\dot{\phi}_{0}}+c_{\phi\gamma}(\dot{\phi}_{0}^{-1}H)^{\cdot}+f_{\phi}+f_{\gamma}\frac{H^{2}}{\dot{\phi}_{0}^{2}}+\tilde{f}_{\phi\gamma}\frac{H}{\dot{\phi}_{0}}+f_{\phi\gamma}\frac{H}{\dot{\phi}_{0}}~.\ee
\be c_{\phi}\equiv\frac{2(1-3\lambda)H^{2}}{d}~,~c_{\gamma}\equiv
\frac{2(1-3\lambda)\dot{\phi}_{0}^{2}}{d}~,~c_{\phi\gamma}\equiv-\frac{4H(1-3\lambda)\dot{\phi}_{0}}{d}~,\ee
\be
f_{\phi}\equiv-\frac{(1-\lambda)}{\kappa^{2}d}(\frac{(1-3\lambda)}{(1-\lambda)}H\dot{\phi}_{0}+V_{0}^{\prime}(\phi_{0}))\dot{\phi}_{0}-\frac{(1-\lambda)}{\kappa^{2}d}V_{4}(\phi_{0})\dot{\phi}_{0}\frac{\partial^{4}}{a^{4}}~\ee
\be f_{\gamma}\equiv
18(1-3\lambda)H-16\kappa^{2}H\frac{(1-3\lambda)}{d}\frac{\partial^{2}}{a^{2}}~\ee
\be f_{\phi\gamma}\equiv
3\kappa^{-2}\dot{\phi}_{0}-\frac{4(1-\lambda)}{d}\dot{\phi}_{0}\frac{\partial^{2}}{a^{2}}~,\ee
\be
\tilde{f}_{\phi\gamma}\equiv+\frac{(1-3\lambda)}{\kappa^{2}d}[\dot{\phi}_{0}^{3}-4H\kappa^{2}V_{0}^{\prime}(\phi_{0})]-\frac{4H}{d}(1-3\lambda)V_{4}(\phi_{0})\frac{\partial^{4}}{a^{4}}~,\ee
\bea
m_{\phi}^{2}&\equiv&-\kappa^{-2}V_{1}(\phi_{0})\frac{\partial^{2}}{a^{2}}-\frac{1}{2\kappa^{2}}\frac{\partial^{2}}{a^{2}}+\kappa^{-2}(V_{2}(\phi_{0})+V_{4}^{\prime}(\phi_{0})\nonumber\\
&+&\frac{(1-\lambda)}{d}V_{4}(\phi_{0})(\frac{(1-3\lambda)}{(1-\lambda)}H\dot{\phi}_{0}+V_{0}^{\prime}(\phi_{0})))\frac{\partial^{4}}{a^{4}}\nonumber\\
&+&\kappa^{-2}V_{6}(\phi_{0})\frac{\partial^{6}}{a^{6}}+\frac{(1-\lambda)}{2\kappa^{2}d}V_{4}^{2}(\phi_{0})\frac{\partial^{8}}{a^{8}}+\frac{1}{2}\kappa^{-2}V_{0}^{\prime\prime}(\phi_{0})\nonumber\\
&+&\frac{\dot{\phi}_{0}^{2}}{4\kappa^{4}}\frac{1}{(1-\lambda)}+\frac{(1-\lambda)}{2\kappa^{2}d}(\frac{(1-3\lambda)}{(1-\lambda)}H\dot{\phi}_{0}+V_{0}^{\prime}(\phi_{0}))^{2}~,\eea
\bea
m_{\gamma}^{2}&\equiv&\frac{\partial^{2}}{a^{2}}+2\kappa^{-2}(8g_{2}+3g_{3})\frac{\partial^{4}}{a^{4}}+\frac{8(1-\lambda)}{d}\kappa^{2}\frac{\partial^{4}}{a^{4}}\nonumber\\
&+&2\kappa^{-4}(8g_{7}-3g_{8})\frac{\partial^{6}}{a^{6}}-9[3(1-3\lambda)H^{2}+\frac{1}{2}\kappa^{-2}\dot{\phi}_{0}^{2}]~,\eea
\bea
m_{\phi\gamma}^{2}&\equiv&3\kappa^{-2}V_{0}^{\prime}(\phi_{0})+\frac{4(1-\lambda)}{d}(\frac{(1-3\lambda)}{(1-\lambda)}H\dot{\phi}_{0}+V_{0}^{\prime}(\phi_{0}))\frac{\partial^{2}}{a^{2}}\nonumber\\
&-&6\kappa^{-2}V_{4}(\phi_{0})\frac{\partial^{4}}{a^{4}}+4\frac{(1-\lambda)}{d}V_{4}(\phi_{0})\frac{\partial^{6}}{a^{6}}~.\eea


\begin{thebibliography}{99}

\bibitem{Guth:1980zm}
  A.~H.~Guth, Phys.\ Rev.\  D {\bf 23}, 347 (1981).

\bibitem{Albrecht:1982wi}
  A.~Albrecht and P.~J.~Steinhardt, Phys.\ Rev.\ Lett.\  {\bf 48},
  1220 (1982);
  A.~D.~Linde, Phys.\ Lett.\  B {\bf 108}, 389 (1982).

\bibitem{Linde:1983gd}
  A.~D.~Linde, Phys.\ Lett.\  B {\bf 129} (1983) 177.

\bibitem{Larson:2010gs}
  D.~Larson {\it et al.},
  Astrophys.\ J.\ Suppl.\  {\bf 192}, 16 (2011).

\bibitem{Stewart:1993bc}
  E.~D.~Stewart and D.~H.~Lyth,
  Phys.\ Lett.\  B {\bf 302}, 171 (1993).

\bibitem{Amsler:2008zzb}
  C.~Amsler {\it et al.}  [Particle Data Group],
  Phys.\ Lett.\  B {\bf 667}, 1 (2008).

\bibitem{CervantesCota:1995tz}
  J.~L.~Cervantes-Cota and H.~Dehnen,
  Nucl.\ Phys.\  B {\bf 442}, 391 (1995).

\bibitem{Bezrukov:2007ep}
  F.~L.~Bezrukov and M.~Shaposhnikov,
  Phys.\ Lett.\  B {\bf 659}, 703 (2008).

\bibitem{Germani:2010gm}
  C.~Germani and A.~Kehagias,
  Phys.\ Rev.\ Lett.\  {\bf 105}, 011302 (2010);
  C.~Germani and A.~Kehagias,
  JCAP {\bf 1005}, 019 (2010)
  [Erratum-ibid.\  {\bf 1006}, E01 (2010)].

\bibitem{Kamada:2010qe}
  K.~Kamada, T.~Kobayashi, M.~Yamaguchi and J.~Yokoyama,
  arXiv:1012.4238 [astro-ph.CO].

\bibitem{Burgess:2009ea}
  C.~P.~Burgess, H.~M.~Lee and M.~Trott,
  JHEP {\bf 0909}, 103 (2009).

\bibitem{Lerner:2009na}
  R.~N.~Lerner and J.~McDonald,
  JCAP {\bf 1004}, 015 (2010);
  J.~L.~F.~Barbon and J.~R.~Espinosa,
  Phys.\ Rev.\  D {\bf 79}, 081302 (2009);
  M.~Atkins and X.~Calmet,
  Phys.\ Lett.\  B {\bf 695}, 298 (2011);
  C.~P.~Burgess, H.~M.~Lee and M.~Trott,
  JHEP {\bf 1007}, 007 (2010);
  M.~P.~Hertzberg,
  JHEP {\bf 1011}, 023 (2010).

\bibitem{Atkins:2010yg}
  M.~Atkins and X.~Calmet,
  Phys.\ Lett.\  B {\bf 697}, 37 (2011).

\bibitem{Nicolis:2008in}
  A.~Nicolis, R.~Rattazzi and E.~Trincherini,
  Phys.\ Rev.\  D {\bf 79}, 064036 (2009).

\bibitem{Horava:2009uw}
  P.~Horava,
  Phys.\ Rev.\  D {\bf 79}, 084008 (2009);
  G.~Calcagni,
  JHEP {\bf 0909}, 112 (2009);
  E.~Kiritsis and G.~Kofinas,
  Nucl.\ Phys.\  B {\bf 821}, 467 (2009).

\bibitem{Mukohyama:2010xz}
  S.~Mukohyama,
  Class.\ Quant.\ Grav.\  {\bf 27}, 223101 (2010);
  T.~P.~Sotiriou,
  J.\ Phys.\ Conf.\ Ser.\  {\bf 283}, 012034 (2011).

\bibitem{Charmousis:2009tc}
  C.~Charmousis, G.~Niz, A.~Padilla and P.~M.~Saffin,
  JHEP {\bf 0908}, 070 (2009);
  M.~Li and Y.~Pang,
  JHEP {\bf 0908}, 015 (2009);
  D.~Blas, O.~Pujolas and S.~Sibiryakov,
  JHEP {\bf 0910}, 029 (2009).

\bibitem{Blas:2009qj}
  D.~Blas, O.~Pujolas and S.~Sibiryakov,
  Phys.\ Rev.\ Lett.\  {\bf 104}, 181302 (2010);
  A.~Papazoglou and T.~P.~Sotiriou,
  Phys.\ Lett.\  B {\bf 685}, 197 (2010);
  D.~Blas, O.~Pujolas and S.~Sibiryakov,
  Phys.\ Lett.\  B {\bf 688}, 350 (2010);
  D.~Blas, O.~Pujolas and S.~Sibiryakov,
  arXiv:1007.3503 [hep-th].

\bibitem{Horava:2010zj}
  P.~Horava and C.~M.~Melby-Thompson,
  Phys.\ Rev.\  D {\bf 82}, 064027 (2010);
  P.~Horava,
  arXiv:1101.1081 [hep-th].

\bibitem{Mukohyama:2009gg}
  S.~Mukohyama,
  JCAP {\bf 0906}, 001 (2009);
  Y.~S.~Piao,
  Phys.\ Lett.\  B {\bf 681}, 1 (2009);
  K.~Izumi, T.~Kobayashi and S.~Mukohyama,
  JCAP {\bf 1010}, 031 (2010).


\bibitem{Gao:2009bx}
  X.~Gao,
  arXiv:0904.4187 [hep-th];
  B.~Chen and Q.~G.~Huang,
  Phys.\ Lett.\  B {\bf 683}, 108 (2010);
  B.~Chen, S.~Pi and J.~Z.~Tang,
  JCAP {\bf 0908}, 007 (2009);
  arXiv:0910.0338 [hep-th];
  X.~Gao, Y.~Wang, R.~Brandenberger and A.~Riotto,
  Phys.\ Rev.\  D {\bf 81}, 083508 (2010);
  Y.~F.~Cai and X.~Zhang,
  Phys.\ Rev.\  D {\bf 80}, 043520 (2009);
  K.~Yamamoto, T.~Kobayashi and G.~Nakamura,
  Phys.\ Rev.\  D {\bf 80}, 063514 (2009);
  A.~Wang and R.~Maartens,
  Phys.\ Rev.\  D {\bf 81}, 024009 (2010);
  JCAP {\bf 1003}, 013 (2010);
  Y.~Lu and Y.~S.~Piao,
  Int.\ J.\ Mod.\ Phys.\  D {\bf 19}, 1905 (2010);
  T.~Kobayashi, Y.~Urakawa and M.~Yamaguchi,
  JCAP {\bf 0911}, 015 (2009);
  JCAP {\bf 1004}, 025 (2010);
  J.~O.~Gong, S.~Koh and M.~Sasaki,
  Phys.\ Rev.\  D {\bf 81}, 084053 (2010);
  A.~Cerioni and R.~H.~Brandenberger,
  arXiv:1007.1006 [hep-th];
  arXiv:1008.3589 [hep-th];
  A.~Wang,
  Phys.\ Rev.\  D {\bf 82}, 124063 (2010).




\end{thebibliography}
\end{document}